\documentclass[final]{svjour2}
\usepackage{graphicx}
\usepackage{rotating}
\usepackage{amssymb}
\usepackage{mathptmx}
\usepackage[numbers]{natbib}
\usepackage{epstopdf}
\def\bra#1{\bigl\langle{ #1} \bigr|}
\def\ket#1{\bigl|{ #1} \bigr\rangle}

\def\ovlp#1#2{\bigl\langle{ #1}\big|{#2} \bigr\rangle}

\def\rvec {{\bf r}}

\def\kvec {{\bf k}}
\def\vec#1{{\bf #1}}
\def\perz#1{\alpha_{#1}^{\dagger}}
\def\pver#1{\alpha_{#1}^{\phantom{\dagger}}}       
\def\qerz#1{a_{#1}^{\dagger}}
\def\qver#1{a_{#1}^{\phantom{\dagger}}}            

\def\EF{e_{\rm F}}
\def\KF{k_{\rm F}}
\def\SF{S_{\rm F}}
\def\a0{a_0}
\def\I{{\rm i}}

\usepackage{soul, xcolor} 
\setstcolor{red}

\makeatletter
\journalname{Journal of Low Temperature Physics}

\bibpunct{}{}{,}{s}{}{,}

\begin{document}

\newcommand{\hdblarrow}{H\makebox[0.9ex][l]{$\downdownarrows$}-}

\title{$^1S_0$ pairing in neutron matter}

\author{H.-H. Fan$^{\dagger\ddagger}$ \and E. Krotscheck$^{\dagger\ddagger}$
  \and J.W. Clark$^{+@}$}
\institute{$^\dagger$Department of Physics, University at Buffalo SUNY,
Buffalo, NY 14260, USA\\
  $^\ddagger$Institut f\"ur Theoretische Physik, Johannes
Kepler Universit\"at, A 4040 Linz, Austria\\
$^+$Department of Physics \& McDonnell Center for the Space Sciences, Washington University, St. Louis MO 63130, USA\\
$^@$ Centro de Ci\^encias Matem\'aticas University of Madeira, 9020-105 
Funchal, Madeira, Portugal}

\maketitle

\begin{abstract}
  
  We report calculations of the superfluid pairing gap in neutron
  matter for the $^1S_0$ components of the Reid soft-core $V_6$ and
  the Argonne $V_{4}'$ two-nucleon interactions. Ground-state
  calculations have been carried out using the central part of the
  operator-basis representation of these interactions to determine
  optimal Jastrow-Feenberg correlations and corresponding effective
  pairing interactions within the correlated-basis formalism (CBF),
  the required matrix elements in the correlated basis being evaluated
  by Fermi hypernetted-chain techniques.  Different implementations of
  the Fermi-Hypernetted Chain Euler-Lagrange method (FHNC-EL) agree at
  the percent level up to nuclear matter saturation density. For the
  assumed interactions, which are realistic within the low density
  range involved in $^1S_0$ neutron pairing, we did not find a
  dimerization instability arising from divergence of the in-medium
  scattering length, as was reported recently for simple square-well
  and Lennard-Jones potential models (Phys. Rev. A {\bf 92}, 023640
  (2015)).
  

\end{abstract}

\section{Introduction} 
\label{sec:intro}
  
\subsection{Adaptation of BCS theory to nuclear systems}
\label{ssec:adaptBCS}

The nature and role of fermionic pairing and superfluidity in nuclei and 
nuclear matter became a subject of great interest shortly after 
publication of the landmark paper by Bardeen, Cooper, and Schrieffer 
(BCS) establishing the physical basis of superconductivity in metals 
\cite{BCS,BCS50book}.  Bohr, Mottelson, and Pines \cite{BMP} 
were quick to recognize implications of this development for a 
deeper understanding of nuclear phenomena, relating it to evidence 
for a characteristic energy gap between the ground state 
and the first intrinsic excitation in a certain class of nuclei.  

Concurrently, there was growing interest among nuclear theorists in 
what could be learned from the quantum many-body problem of infinite 
nuclear matter composed of nucleons interacting through the best 
nucleon-nucleon (NN) potentials available at the time.  Cooper, Mills, 
and Sessler\cite{CMS} (CMS) were the first to apply BCS theory to 
such a system.  They encountered two obstacles when attempting to 
solve the BCS equation for the superfluid energy gap $\Delta_{\kvec}$ 
as a function of momentum ${\kvec}$.

To understand what they faced, it is necessary to consider the
BCS gap equation, written in the generic form
\begin{equation}
\Delta_{\kvec} = - \sum_{{\kvec}'}
P_{\kvec,{\kvec}'} {\Delta_{{\kvec}'}\over{2 E_{{\kvec}'}}}\,,
\label{eq:geq}
\end{equation}
where $P_{{\kvec},{\kvec}'} = \langle {\kvec} \uparrow,-{\kvec} \downarrow 
|v(12)| {\kvec}' \uparrow , - {\kvec}' \downarrow \rangle $ defines the 
pairing matrix elements of the bare two-body potential $v(12)$, while
\begin{equation}
E_{\kvec} = [(e_{\kvec}-\mu)^2 + \Delta^2_{\kvec}]^{1/2} \label{eq:eden}
\end{equation}
represents the (gapped) quasiparticle energy in the superfluid state,
with $e_{\kvec}$ an ``appropriate'' single-particle energy related
to the normal state.  Given the original BCS trial ground state
\begin{equation}
\left|{\rm BCS} \right\rangle = 
{\prod_{\kvec}}
\left[ u_{\kvec} +  v_{\kvec} a_{ {\bf  k} \uparrow }^\dagger
 a_{-{\bf  k} \downarrow}^\dagger  \right] |0 \rangle
\label{eq:BCS} 
\end{equation}
(but written slightly differently in terms of Bogoliubov amplitudes 
$u_{\kvec}$, $v_{\kvec}$ satisfying the normalizing condition 
$u_{\kvec}^2 + v_{\kvec}^2 = 1$), the expression (\ref{eq:geq}) of the 
gap equation can be derived from the Euler-Lagrange variational 
principle following exactly the same path as in the 1957 BCS paper 
\cite{BCS} and in Schrieffer's book \cite{Schrieffer1999}.  As the BCS 
state does not have a definite particle number, the chemical potential 
$\mu$ (determined from the number density) is introduced as a Lagrange 
parameter to accommodate the constraint that the particle number is 
conserved on average.

Of the two problems CMS faced in implementing BCS theory for nuclear 
matter, they managed to solve what appeared to be the more difficult one, 
and finessed the other.  During this same period in the mid-to-late 
1950s, it had become apparent that an acceptable model of the NN 
interaction, fitted to the available NN scattering data and the deuteron, 
must possess a strong inner repulsion, most commonly taken to be a hard core.  
This precluded solving the BCS gap equation as formulated in momentum 
space, because the necessary pairing matrix elements $P_{{\kvec},{\kvec}'}$ 
of the NN potential would be undefined.  However, CMS recognized that 
the BCS gap equation could be transformed to coordinate space to yield
a nonlinear but Schr\"odinger-like equation for an underlying two-body 
problem.  The analog of the wave function for the separation vector 
${\bf r}$ is the pairing function $\chi({\bf r})$, which may also 
be regarded as the superfluid order parameter. Basically, $\chi({\bf r})$ 
is the Fourier transform of the product $u_{\kvec}v_{\kvec}$ 
of Bogoliubov amplitudes, or equivalently of $\Delta({\kvec})/2E_{\kvec}$. 
Therefore the problem created by the hard core of the NN potential could 
be solved, for the same reason that the Schr\"odinger equation for a 
hard-sphere scattering potential has a solution.  

The second problem confronting CMS was what to take for the single-particle 
energy $e_{\kvec}$ in the expression for $E_{\kvec}$.  There is first 
a subtlety relating to $e_{\kvec}$ that should be exposed, for the record.
The above derivation leads to the actual expression

\begin{equation}
e_{\kvec} = { {\hbar^2 k^2 } \over {2m} }
 + \frac{1}{2} \sum_{{\bf l} \sigma \sigma^\prime } v_{\bf l}^2
 \langle {\kvec} \sigma , {\bf l} \sigma^\prime
                | v(12) | {\kvec} \sigma , {\bf l} \sigma^\prime
               - {\bf l} \sigma^\prime , {\kvec} \sigma \rangle\,.
\label{eq:speeq}
\end{equation}
This contains the Fermi-surface smearing factor represented by 
$v_{\bf l}^2$, and hence requires a solution of the pairing problem before
$e_{\kvec}$ can be evaluated.  In practice, this factor is almost
always replaced by the Fermi step, converting $e_{\kvec}$ to a
standard Hartree-Fock single-particle energy.  It is argued, in most
cases safely, that this can be done because the gap $\Delta_{\kvec}$
is much smaller than the Fermi energy, thus decoupling $e_{\kvec}$
from the rest of the gap problem.

The primary issue raised by the expression (\ref{eq:speeq}) is not at all
subtle.  If the bare NN interaction contains a hard core, the Hartree-Fock
matrix elements it contains are infinite; nor would the results for
$e_{\kvec}$ be sensible if the interaction remains finite, but features 
an internal repulsion strong enough to achieve empirical saturation
of nuclear matter. CMS were forced to finesse this second problem; they 
imposed an effective-mass spectrum $e_{\kvec} = \hbar^2k^2/2m^*$.  With 
this step, the problem was well-defined and in principal soluble 
for $\Delta_{\kvec}$; however, for a time only the {\it existence\/} of 
a superfluid solution was established \cite{Mills}, due to the limited 
computational resources of that period. 

In summary, the nature of the BCS theory of superfluidity is such that
its application to nuclear systems is practical, in particular for the
hypothetical system of infinite nuclear matter and certain nucleonic
subsystems existing in neutron stars.  However, due to the presence of
a strong short-range repulsion in the bare NN interaction, one must
make a reasonable, but {\it ad hoc\/}, assumption for the normal-state
single-particle energy.  The theory has the capacity to generate
two-body correlations that can accommodate even the effects of a hard
core, although the problem must then be solved in coordinate space.
Solution of the problem in momentum space, {\em i.e.}\ the original
gap equation (\ref{eq:geq}), does in fact become possible if the NN
interaction, even though strongly repulsive at short distance, has a
Fourier transform.  (For some interactions including the Reid
soft-core potential \cite{Reid68,JWCgap}, numerical solution can
present some technical difficulty; this can be avoided by applying the
separation approach developed in Ref.~\citenum{KKC96}).

Yet the status of nuclear BCS as described remains unsatisfactory for 
potentials with repulsive cores.  This issue was naturally addressed 
by the introduction of Jastrow-Feenberg correlation factors 
\cite{YangNC,YangClarkBCS,YangThesis}.
Cluster-expansion techniques were applied to evaluate the required
expectation values\cite{Johnreview}.  The corresponding gap equations
were studied and procedures for their solution explored, with applications 
not only to isospin-symmetric nuclear matter (equal numbers of
neutrons and protons) \cite{YangNC}, but also pure neutron matter and
$\beta$-stable nucleonic matter relevant to neutron-star interiors
\cite{YangClarkBCS,Chao}.  In the mid-1970s, major advances in
microscopic quantum many-body theory involving correlated basis
functions (CBF) were made with the replacement of cluster expansions
by Fermi Hypernetted-Chain (FHNC) diagram-resummation techniques
\cite{Johnreview,KroTrieste}, facilitating accurate evaluation of
expectation values and matrix elements of observables in a correlated
basis and culminating in a framework for Euler-Lagrange optimization
of Jastrow-Feenberg correlations.  When implemented in a BCS
extension, these advances have made possible the development of a
rigorous correlated BCS (CBCS) theory \cite{CBFPairing} (see also
Ref.~\citenum{HNCBCS}) that respects the U(1) symmetry-breaking aspect
of the superfluid state -- {\em i.e.}, non-conservation of particle
number.  Some earlier applications of CBCS theory to nuclear systems,
and especially neutron-star matter, may be found in
Refs.~\citenum{JWCgap,CBFPairing}.  A recent in-depth study of
correlations in the low-density Fermi gas\cite{cbcs}, with emphasis on
the presence of Cooper pairing and dimerization, documents the power
of the Euler-Lagrange FHNC approach adopted in the present work, especially 
when coupled with CBF perturbation theory.

\subsection{Extension of BCS asymptotics to structured interactions
and in-medium effects} 
\label{ssec:asymptotics}

Having derived the equivalent of the gap equation (\ref{eq:geq}), BCS 
went on to simplify the pairing interaction in a way suitable for
electron liquids in solids, arriving at the iconic asymptotic result
\begin{equation}
\Delta \simeq 2 \hbar \omega_c e^{-1/\lambda} 
\label{eq:asym}
\end{equation}
for the value of the energy gap $\Delta$ in terms of a cutoff
$\hbar \omega_c$ and the coupling constant $\lambda = |N(0)V|$
of the attractive pairing interaction $V$, with $N(0)$ denoting 
the density of states around the Fermi surface. It is important to 
recognize that this result, being restricted to the weak-coupling
regime $\lambda \ll 1$, is not at all appropriate for nuclear problems. 
In nuclear systems, the bare two-body interaction is strong, and
strongly non-monotonic in coordinate space.  Two parameters are 
not sufficient to characterize the asymptotic behavior of the gap at 
relevant densities. See Refs.~\citenum{KKC96,mag7} for extensive 
analysis and computational exploration of this important distinction.  
The latter reference includes an asymptotic study in which the pairing 
interaction is characterized by an additional parameter $\kappa$ 
along with the traditional coupling constant $\lambda$ and cutoff 
frequency $\omega$.  This ``stiffness'' parameter is introduced to 
represent a nontrivial momentum dependence of the pairing interaction 
$P_{{\kvec}{\kvec'}}$.  Asymptotic behavior in the four quadrants 
$(\lambda \pm $,\, $\kappa \pm)$ is explored in Ref.~\citenum{mag7}, 
pointing to the existence of solutions with behavior quite distinct 
from the familiar relation (\ref{eq:asym}), in addition to a BCS-analog.

Another asymptotic formula of special interest (and of long standing) is
that of Gorkov and Melik-Barkhudarov \cite{Gorkov} (GM), 
\begin{equation}
\Delta_F = (4e)^{-1/3}\frac{8}{e^2} e_F e^{-1/\lambda}\,, 
\end{equation}
written for the zero-temperature gap rather than the critical
temperature.  This result was derived by field-theoretic methods in
the limit of an infinitely dilute gas of interacting spin-${1/2}$
fermions, with $\lambda = 2k_F|a_0|/\pi$.  Here $a_0$ is the {\it
  vacuum} scattering length, assumed to be negative, $e_F = \hbar^2
\KF^2/2m$ is the Fermi energy, $m$ the fermion mass, and $\KF$ the Fermi
momentum.  The prefactor $(4e)^{-1/3}$ is an
in-medium correction for a polarization-induced interaction
corresponding to exchange of virtual phonons.  The same result without
the GM prefactor was re-derived several times in the 1990's, basically
by summing ladder diagrams for the bare interaction (see
Ref.~\citenum{heiselbergPRL00}, where the GM prefactor is generalized
to $(4e)^{\nu/3-1}$ for an arbitrary number $\nu$ of fermion species).

In the recent work previously cited\cite{cbcs}, it has been argued
(cf.~Eqs.~(3.25) and (3.26)) that if one has corrections of the
in-medium scattering length $a$ to the vacuum scattering length of the
form
\begin{equation}
a  = \a0\left[1 + \alpha\frac{\a0\KF}{\pi}+\ldots\right]\,,
\end{equation}
it follows that
\begin{equation}
\Delta_F =
\frac{8}{e^2}\EF\exp\left(-\frac{\alpha}{2}\right)
\exp\left(\frac{\pi}{2\a0\KF}\right)\,.
\end{equation}
The GM factor is just one of these corrections, which still assumes
that the pairing matrix element at $\KF$ is the same as that at
$k=0$. Removing this assumption produces yet another correction of the
same kind.

The above summary of BCS asymptotics is intended to provide deep 
background for the present work on neutron matter at densities occurring 
in the inner-crust region of neutron stars, but their direct relevance
is open to question.  The neutron densities involved in this application 
are low compared to the saturation density $\rho_0$ of isospin-symmetric 
nuclear matter, which, in pure neutron matter, would correspond to a $\KF$ 
value of about $1.7~{\rm fm}^{-1}$.   We will find that $^1S_0$ pairing in 
neutron matter is strongest at somewhat less than half that value, thus at a 
density an order of magnitude below $\rho_0$.  On the other hand, given the 
unusually large magnitude of the neutron-neutron $S$-wave scattering length, 
$\a0 \approx -18.6~{\rm fm}$, the diluteness condition $|\a0|\KF \ll 1$ 
implies $\KF \ll 0.05~{\rm fm}^{-1}$, over three orders of magnitude lower in 
density than that of the physically relevant neutron-star environment.
Naturally the dilute-limit asymptotics do apply for the extreme 
low-density tail of the roughly Gaussian shape of $\Delta_F$
vs.\ $\KF$ in the $^1S_0$ neutron pairing problem considered here.  
The higher-density tail is more relevant; it has been demonstrated in
Ref.~\citenum{KKC96} that $\Delta_F$ dies exponentially to naught 
as an upper critical density is approached.

\subsection{Sensitivity issues in optimization}
\label{ssec:sensitivity}

The foregoing subsections of this introduction provide a rather
elaborate background and motivation for the work to be presented.
Another motivation is more immediate.  Recently, using updated 
modern NN interactions, gap calculations \cite{Pavlou2017} for 
pure neutron matter have again been performed within the simpler 
version of correlated BCS theory in which the ground-state energy 
for a Jastrow-Feenberg trial function, estimated by a truncated 
cluster expansion, is minimized with respect to the parameters in 
an assumed analytic form for the Jastrow two-body correlation function 
$f(r)$.  The correlation function so determined is used to generate 
a ``tamed'' effective pairing interaction for calculation of a 
corresponding superfluid gap in the $^1S_0$ state.  

Ref.~\citenum{Pavlou2017} presents results for the ground-state energy 
per particle $E/N$ and the corresponding $^1S_0$ energy gap, based on 
the Argonne $V_{18}$ (AV18) NN interaction\cite{AV18} and two trial correlation 
functions with analytic forms that have been employed in earlier 
Jastrow-Feenberg studies of nuclear and neutron matter.  The optimal 
ground-state energies determined for these two choices show only minor 
quantitative differences over the low range of densities where a 
significant $^1S_0$ gap is to be expected (peaking at about 1/8 nuclear 
saturation density).  The two curves obtained for the gap $\Delta_{\rm F}
=\Delta(k=\KF)$ at the Fermi surface, plotted versus Fermi momentum 
$\KF$, have a Gaussian appearance.  In contrast to the close agreement 
of the $E/N$ results for the two correlation choices, the corresponding
peak values for $\Delta_{\rm F}$ are found to differ by almost a factor 
two (with a value 1.8 MeV for the correlation function featuring an 
overshoot of unity versus 3.3 MeV for one that does not.

This finding could be interpreted as a reflection of the variational
property that a small error of order $\delta$ in the wave function
only entails an error of order $\delta^2$ in the energy expectation
value, but of order $\delta$ for other observables, with $\delta$ in
this case corresponding to the difference in the choices for $f(r)$.
But the situation may actually be worse for two reasons: The most
immediate one is that the gap itself shows an exponential
amplification of errors in the coupling strength and density of
states, at least for the standard BCS case.  Some results of the
present investigation indicate a similar strong sensitivity of gap
behavior. The second, more subtle reason, involves the convergence of
cluster expansions for correlated wave functions: Typically, the
contribution of an $n$-body diagram in the energy is amplified by a
factor of $n^2$ in its contribution to the effective interactions
needed to calculate the coupling matrix elements.

Figure 6 of the paper of Pavlou {\em et al.}~\cite{Pavlou2017} shows plots
of $\Delta_F$ versus $\KF$ for the AV18 interaction as obtained in 
almost a dozen calculations by different theoretical methods, including 
various versions of Monte Carlo.  (Actually, this is a summary figure 
taken from the review by Gezerlis {\em et al.}~\cite{Gezerlis2014} of novel 
superfluidity in neutron stars, with a curve calculated by Pavlou 
{\em et al.\/} superimposed.) There is a spread of a factor of six in 
the peak values of $\Delta_F$, with a significant spread also in 
the peak densities.  In view of what has been said above, this is 
hardly surprising, although the calculations may differ in the 
inclusion of in-medium effects.

At any rate, the message from the considerations of this subsection is
that it is imperative within any variational approach to seek truly
optimal correlations, without resorting to simple parametrizations, 
and that is what Euler-Lagrange FHNC (FHNC-EL) can deliver, with 
minimal error.

The rest of this paper is organized as follows.  Section~\ref{sec:GMBT}
exemplifies what qualifies as a generic many-body theory, first with a
brief review of the elements of the Jastrow-Feenberg theory of the normal 
ground state of a many-fermion system (Sec.~\ref{ssec:FHNC}), then with an
introduction to the formalism associated with the method of correlated
basis functions (CBF) (Sec.~\ref{ssec:CBF}), concluding with the essentials 
of a coherent theory of fermion superfluidity within the CBF framework 
(Sec.~\ref{ssec:CBCS}), based on Euler-Lagrange Fermi Hypernetted-Chain    
optimization (FHNC-EL).  In Sec.~\ref{ssec:energetics}, we describe and 
discuss our application of two types of FHNC-EL theory to the 
ground-state energetics of pure neutron matter.  Sec.~\ref{ssec:bcs} 
is concerned with solution of the resulting CBF gap equation for $^1S_0$ 
pairing, which incorporates the effects of the optimal Jastrow-Feenberg 
correlations.  Results for energetics (the equation of state) and BCS
pairing in CBF framework are presented and discussed for two versions of 
the bare NN interaction, namely the Reid soft-core $V_6$ potential 
\cite{Reid68} and the Argonne $V_4'$ interaction\cite{Wiringa2002}.  
Well known from earlier microscopic studies of nuclear matter, these 
choices are quantitatively viable in the low-density regime where 
the $^1S_0$ pairing state is dominant.  Only the central components of 
these potentials and their $^1S_0$ projections are needed for determination 
of the CBF pairing matrix elements, in contrast to the case of the full 
Argonne $V_{18}$ interaction \cite{AV18}.  (Note that Argonne $V_4'$ is
central by construction and we use only the $V_4$ part of the Reid potential,
{\em i.e.,\/} its tensor as well as spin-orbit terms being omitted.)  Many-body
aspects of our findings unique to optimal incorporation of short- and
long-range correlations within the CBF/FHNC formalism are analyzed.
Where meaningful, our predictions for the density dependence of the 
gap at the Fermi wave number are compared with those from other microscopic 
calculations.  Sec.~\ref{sec:summary} summarizes ways in which the 
present numerical study may be improved and extended.

\section{Generic Many-Body Theory}
\label{sec:GMBT}
\subsection{The normal ground state}
\label{ssec:FHNC}

In this section, we briefly describe the Jastrow-Feenberg variational
method and its implementation to superfluid systems.  (For comprehensive 
background on this many-body approach and its generalization to the method 
of correlated basis functions, see Refs.~\citenum{Johnreview,KroTrieste}.  
Recent descriptions and analysis of its applications to superfluid 
systems may be found in Refs.~\citenum{cbcs} and \citenum{50yrs}).  We 
call this method ``generic many-body theory'' because the same equations 
can be derived by Green functions methods \cite{parquet1,parquet2}, from
coupled-cluster theory \cite{BishopValencia}, and by a generalization
of density functional theory to pair distribution functions
\cite{PairDFT}, without mentioning a Jastrow-Feenberg wave function. 
We use the Jastrow-Feenberg point of view here because it is
the simplest to implement and to generalize.

For a strongly interacting and translationally invariant {\em
normal\/} system, the Jastrow-Feenberg theory assumes a
non-relativistic many-body Hamiltonian
\begin{equation}
H = -\sum_{i}\frac{\hbar^2}{2m}\nabla_i^2 + \sum_{i<j}
v(i,j)\,.
\label{eq:Hamiltonian}
\end{equation}
The method starts with an {\em ansatz\/} for the wave function,
\cite{FeenbergBook}
\begin{eqnarray}
\Psi_0({\bf r}_1,\ldots,{\bf r}_N) &=& \frac{1}{\sqrt{I_{{\bf o},{\bf o}}}}
	F({\bf r}_1,\ldots,{\bf r}_N)
	\Phi_0({\bf r}_1,\ldots,{\bf r}_N)\label{eq:wavefunction},\\
F({\bf r}_1,\ldots,{\bf r}_N) &=& \exp\frac{1}{2}
\left[\sum_{i<j}  u_2({\bf r}_i,{\bf r}_j) + \cdot\cdot +  
\sum_{i_1<\ldots<i_n}u_n({\bf r}_{i_1},.., {\bf r}_{i_n}) 
+ \cdot\cdot \right]\,,
\label{eq:Jastrow}
\end{eqnarray}
where
${I_{{\bf o},{\bf o}}} = \left\langle \Phi_0 | F^\dagger F |
\Phi_0\right\rangle$ is a normalizing constant. 
Here $\Phi_0({\bf r}_1,\ldots,{\bf r}_N)$ denotes a model state, 
which for normal Fermi systems is a Slater-de\-ter\-mi\-nant, and $F$ 
is a correlation operator written in general form, but to be truncated 
at the two-body $u_2$ term in a standard Jastrow calculation.  There are 
basically two ways to deal with this type of wave function. In quantum 
Monte Carlo studies, the wave function (\ref{eq:wavefunction}) is 
referred to as ``fixed-node approximation,'' and an optimal correlation 
function $F({\bf r}_1,\ldots,{\bf r}_N)$ is obtained by stochastic
means. Computationally far less demanding are diagrammatic methods,
specifically the optimized Euler-Lagrange Fermi-hypernetted chain
(FHNC-EL) method, which is well suited for calculation of physically
interesting quantities.  These diagrammatic methods have been
successfully applied to such highly correlated Fermi systems as $^3$He
at $T=0$~\cite{polish}. We have shown in recent work \cite{ljium} that
even the simplest version of the FHNC-EL theory is accurate within
better than one percent at densities less than 25\% of the saturation
density of liquid $^3$He, and the same or better performance is 
expected for nuclear systems.

The correlations $u_n({\bf r}_1,\ldots,{\bf r}_n)$ are obtained by minimizing
the energy, {\em  i.e.\/} by solving the Euler-Lagrange (EL) equations
\begin{eqnarray}
&&E_0 = \left\langle\Psi_0\right|H\left|\Psi_0\right\rangle
\equiv H_{{\bf o},{\bf o}}\label{eq:energy}\,,\\
&&        \frac{\delta E_0}
{\delta u_n}({\bf r}_1,\ldots,{\bf r}_n) = 0\,.
\label{eq:euler}
\end{eqnarray}

Evaluation of the energy (\ref{eq:energy}) for the variational wave
function (\ref{eq:wavefunction}), (\ref{eq:Jastrow}) and analysis of the
variational problem are carried out by cluster expansion and
resummation methods.  The procedure has been described at length in
review articles \cite{Johnreview,polish} and pedagogical material
\cite{KroTrieste}.  Here, we spell out the simplest version of the
equations that is consistent with the variational problem
(``FHNC//0-EL''). These equations do not provide the quantitatively
best implementation of this approach \cite{polish}. Instead, they
provide the {\em minimal\/} version of the FHNC-EL theory. In
particular, they contain the indispensable physics, namely the correct
description of both short- and long-ranged correlations.

In the FHNC//0-EL approximation, which contains both the random phase
approximation (RPA) and the Bethe-Goldstone equation in a ``collective'' 
approximation, the Euler equation (\ref{eq:euler}) can be written 
in the form
\begin{equation}
S(k) = \frac{\SF(k)}{\sqrt{1 +
	2\frac{\displaystyle \SF^2(k)}{\displaystyle t(k)}
\tilde V_{\rm p-h}(k)}}\,,
\label{eq:FermiRPA0}
\end{equation}
where $S(k)$ is the static structure factor of the interacting system,
$t(k) = \hbar^2 k^2/2m$ is the kinetic energy of a free particle,
$\SF(k)$ is the static structure factor of the non-interacting Fermi 
system, and
\begin{equation}
V_{\rm p-h}(r) =
\left[1+ \Gamma_{\!\rm dd}(r)\right]v(r)
+ \frac{\hbar^2}{m}\left|\nabla\sqrt{1+\Gamma_{\!\rm dd}(r)}\right|^2
+ \Gamma_{\!\rm dd}(r)w_{\rm I}(r)
\label{eq:VddFermi0}
\end{equation}
is the so-called ``particle-hole interaction.''  As usual, we define
the Fourier transform with a density factor, {\em i.e.},
\begin{equation}
\tilde f(\kvec) \equiv \rho \int d^3r\, e^{\I\kvec\cdot\rvec} f(\rvec)\,.
\label{eq:Fouri}
\end{equation}
Auxiliary quantities are the ``induced interaction''
\begin{equation}
\tilde w_{\rm I}(k)=-t(k)
\left[\frac{1}{\SF(k)}-\frac{1}{ S(k)}\right]^2
\left[\frac{S(k)}{\SF(k)}+\frac{1}{2}\right]
\label{eq:inducedFermi0}
\end{equation}
and the ``direct-direct correlation function,'' 
\begin{equation}
{\widetilde \Gamma}_{\!\rm dd}(k) = \bigl(S(k)-\SF(k)\bigr)/\SF^2(k)\,,
\label{eq:GFHNC}
\end{equation}
a ``dressed'' analog of the Fourier inverse of $\exp[u_2(r)] -1$.
Eqs.~(\ref{eq:FermiRPA0})-(\ref{eq:GFHNC}) form a closed set which
can be solved by iteration.  Note that the Jastrow correlation
function $f(r) = \exp(u(r)/2)$ has been eliminated entirely.

More complicated versions of the FHNC-EL method add additional
equations for the so-called ``ee'', ``de,'' and ``cc'' diagrams
which have been expressed in detail in Refs.~\citenum{annals} and
\citenum{polish}; they will not be repeated here.

\subsection{Correlated Basis Functions}
\label{ssec:CBF}

Correlated Basis Functions (CBF) theory uses the correlation operator
$F$ to generate a complete set of basis states through
\begin{equation}
\vert \Psi_{\bf m}^{(N)} \rangle =
\frac{F_{\!N} \; \vert \Phi_{\bf m}^{(N)} \rangle }
{\langle \Phi_{\bf m}^{(N)} \vert  F_{\!N}^{\dagger} F^{\phantom{\dagger}}_{\!N}
\vert \Phi_{\bf m}^{(N)}
\rangle^{1/2} } \,,
\label{eq:States}
\end{equation}
where the $\{\vert \Phi_{\bf m}^{(N)} \rangle\}$ are Slater
determinants of single-particle orbitals.  We review the method only
very briefly, the diagrammatic construction of the relevant
ingredients having been derived in Ref.~\citenum{CBF2} (see also
Ref.~\citenum{polish} for further details).

To develop a BCS theory with correlated wave functions it is
expedient to introduce a ``second-quantized'' formulation.  The
Jastrow-Feenberg correlation operator in (\ref{eq:Jastrow}) depends
on the particle number, {\it i.e.\/} $F=F_{\!N}(1,\ldots,N)$ (whenever
unambiguous, we omit the corresponding subscript). Starting from the
conventional $\qerz{k},\, \qver{k}$ operators that create and annihilate 
single-particle states, new creation and annihilation operators
$\perz{k},\, \pver{k}$ of {\em correlated states\/} are defined by their
action on the correlated basis states:
\begin{eqnarray}
\perz{k}\,\bigl|\Psi_{\bf m}\bigr\rangle
&\equiv\>& \frac{ F_{\!\!_{N+1}} \qerz{k} \,\ket {\Phi_{\bf m}} }{
\bra {\Phi_{\bf m}} \qver{k} F_{\!\!_{N+1}}^\dagger 
 F_{\!\!_{N+1}}^{\phantom{\dagger}}
 \qerz{k}\ket {\Phi_{\bf m}}^{1/2} }\, ,
\label{eq:creation}\\
\pver{k}\,\bigl|\Psi_{\bf m}\bigr\rangle
&\equiv\>& \frac{ F_{\!\!_{N-1}} \qver{k}\,\ket {\Phi_{\bf m}} }{
\bra {\Phi_{\bf m}} \qerz{k} F_{\!\!_{N-1}}^\dagger 
F_{\!\!_{N-1}}^{\phantom{\dagger}}
a_{k}\ket {\Phi_{\bf m}}^{1/2} }\,.
\label{eq:annihilation}\end{eqnarray}
According to these definitions, $\alpha_{k}^\dagger$ and
$\alpha^{\phantom{\dagger}}_{k}$ obey the same commutation rules as
the creation and annihilation operators $\qerz{k}$ and $\qver{k}$ of
uncorrelated states, but they are not Hermitian conjugates of one
another.

For off-diagonal elements $O_{\bf m,n}$ of an $n$-body operator $O$, we
sort the quantum numbers $m_i$ and $n_i$ such that $|\Psi_{\bf m}
\rangle$ is mapped onto $\left|\Psi_{\bf n}\right\rangle$ by
\begin{equation}
\label{eq:defwave}
\left|\Psi_{\bf m}\right\rangle = \perz{m_1}\perz{m_2}
\cdots 
\perz{m_d} \; \pver{n_d} \cdots \pver{n_2}\pver{n_1}  
\left|\Psi_{\bf n} \right\rangle \,.
\end{equation}
Then, the matrix elements $O_{\bf m,n}$ depend only on the 
{\it difference\/} 
between the states
$\vert \Psi_{\bf m} \rangle$ and $\vert \Psi_{\bf n} \rangle$, and {\it
not\/} on the states as a whole.  Consequently, $O_{\bf m,n}$ can be
written as the matrix element of a $d$-body operator
\begin{equation}
\label{eq:defmatrix}
O_{\bf m,n} = \bra{\Psi_{\bf m}} O\ket{ \Psi_{\bf n}}
\equiv \bra{ m_1\, m_2 \, \ldots m_d \,}
{\cal O}(1,2,\ldots d) \,\ket{n_1\,
n_2 \, \ldots n_d}_a \,,
\end{equation}
with the index $a$ indicating antisymmetrization.  In homogeneous systems,
the continuous parts of the quantum numbers $m_i,\,n_i$ are wave
numbers ${\bf p}_i,\,{\bf p}'_i$; we abbreviate their difference as
${\bf q}_i$.

The key quantities for the execution of the theory are diagonal and 
off-diagonal matrix elements of unity and $H'\!\equiv H\!-
\!H_{{\bf o},{\bf o}}$,
\begin{eqnarray}
M_{\bf m,n} &=& \ovlp{\Psi_{\bf m}}{ \Psi_{\bf n}}
\equiv \delta_{\bf m,n} +  N_{\bf m,n}\;,
\label{eq:defineNM}
\\
H'_{\bf m,n} &\equiv &
W_{\bf m,n} + \frac{1}{2}\left(H'_{\bf m,m}+H'_{\bf n,n}\right)N_{\bf m,n} \,.
\label{eq:defineW}
\end{eqnarray}
Eq. (\ref{eq:defineW}) defines a natural decomposition \cite{KroTrieste,CBF2} 
of the matrix elements of $H'_{\bf m,n}$ into the off-diagonal quantities 
$W_{\bf m,n}$ and $N_{\bf m,n}$ and diagonal quantities $H'_{\bf m,m}$. 
These diagonal matrix elements are additive to leading order in the particle
number, allowing us to define the CBF single-particle energies $e_{m}$
that satisfy
\begin{equation}
H'_{\bf m,m} = \bra{\Psi_{\bf m}} H' \ket{\Psi_{\bf m}} \>\equiv\>
\sum_{i=1}^d\left[e_{m_i}-e_{n_i}\right]  + {\cal O}(N^{-1}) \,.
\label{eq:CBFph}
\end{equation}

According to Eq.~(\ref{eq:defmatrix}),
$W_{{\bf m},{\bf n}}$  and $N_{{\bf m},{\bf n}}$ define 
$d$-particle operators ${\cal N}$ and ${\cal W}$, {\em e.g.\/}
\begin{eqnarray}
N_{{\bf m},{\bf o}} &\equiv& N_{p_1p_2\ldots p_d \,h_1h_2\ldots h_d,0} \nonumber\\
&\equiv& \langle p_1p_2\ldots p_d \,|\, {\cal N}(1,2,\ldots,d)\,
|\,h_1h_2\ldots h_d \rangle_a  \;,\nonumber\\
W_{{\bf m},{\bf o}} &\equiv& W_{p_1p_2\ldots p_d \,h_1h_2\ldots h_d,0}\nonumber\\
&\equiv&  \langle p_1p_2\ldots p_d \,|\, {\cal W}(1,2,\ldots,d)\,
|\,h_1h_2\ldots h_d \rangle_a  \,.
\label{eq:NWop}
\end{eqnarray}
Diagrammatic representations of ${\cal N}(1,2,\ldots,d)$ and ${\cal
W}(1,2,\ldots,d)$ have the same topology \cite{CBF2}.  In the next
section, we will show that in dealing with pairing phenomena, 
only the two-body operators are needed.

In principle, ${\cal N}(1,2)$ and ${\cal W}(1,2)$ are non-local
$2$-body operators.  The leading, local contributions to these
operators are readily expressed in terms of the diagrammatic
quantities of FHNC-EL theory \cite{polish}:
\begin{eqnarray}
{\cal N}(1,2) &=& {\cal N}(r_{12})\, =\, \Gamma_{\rm dd}(r_{12})\,,\nonumber\\
{\cal W}(1,2) &=& {\cal W}(r_{12})\,,\quad \tilde {\cal W}(k) =
- \frac{t(k)}{\SF(k)}\tilde \Gamma_{\rm dd}(k)\,.
\label{eq:NWloc}
\end{eqnarray}
For further reference we also display the coordinate space
form of the interaction $ {\cal W}(r_{12})$:
\begin{eqnarray}
  {\cal W}(r) &=&  V_{\rm p-h}(r) + w_{\rm I}(r),\nonumber\\
  &=& \left[1+\Gamma_{\rm dd}(r)\right](v(r)+w_{\rm I}(r)) + 
\frac{\hbar^2}{m}\left|\nabla\sqrt{1+\Gamma_{\!\rm dd}(r)}\right|^2\,,
\label{eq:Weff}
\end{eqnarray}
which exhibits somewhat more clearly the physical meaning of the
individual terms: The factor $\left[1+\Gamma_{\rm dd}(r)\right]$
describes the short-ranged correlations, the term
$(\hbar^2/m)|\nabla\sqrt{1+\Gamma_{\!\rm dd}(r)}|^2$
describes the cost in kinetic energy for bending the wave functions at
short distances, and the induced potential $w_{\rm I}(r)$ describes the
corrections due to phonon exchange. In the local approximations
spelled out in Eqs. (\ref{eq:NWloc}), the CBF single-particle
energies (\ref{eq:CBFph}) assume the simple form
\begin{equation}
  e_k = t(k) + \frac{\tilde X'_{\rm cc}(k)}{1-\tilde X_{\rm cc}(k)}
  + {\rm const.}\label{eq:CBFspect}
\end{equation}
with
\begin{eqnarray}
  \tilde X'_{\rm cc}(k) &=& -\frac{\rho}{\nu}
  \int d^3r\, e^{\I\kvec\cdot\rvec}{\cal W}(r)\ell(r\KF)\,,\\
  \tilde X_{\rm cc}(k) &=& -\frac{\rho}{\nu}
  \int d^3r\, e^{\I\kvec\cdot\rvec}\Gamma_{\rm dd}(r)\ell(r\KF)\,,
\end{eqnarray}
where $\nu\ (=2)$ is the degree of degeneracy of the single-particle
states, $\ell(x) = (3/x)j_1(x)$ is the Slater exchange function,
and the constant is determined by the condition $e_{\KF} = \mu$.
In the limit of a weakly interacting system, we have ${\cal W}(r)
= v(r)$, and the $e_k$ reduce to the Hartree-Fock single-particle
energies (\ref{eq:speeq}).

\subsection{BCS Theory with correlated wave functions}
\label{ssec:CBCS}

The BCS theory of fermion superfluidity ge\-ne\-ra\-li\-zes the Hartree-Fock 
model by introducing a superposition of independent-particle wave 
functions corresponding to different particle numbers 
\cite{BeliaevLesHouches}, represented economically by Eq.~(\ref{eq:BCS})
in terms of Bogoliubov amplitudes $u_{\vec k}$, $v_{\vec k}$.
 
The most natural way to deal with a strongly correlated system is to
first project the bare BCS state on an arbitrary member of a complete
set of independent-particle states with fixed particle numbers.
Then apply the correlation operator to that state, normalize the result, and
finally, sum over all particle numbers $N$. Thus, the correlated BCS
(CBCS) state is taken as
\begin{equation}
\ket{\rm CBCS} =  \sum_{{\bf m},N} \ket {\Psi_{\bf m}^{(N)}}
\langle\Phi_{\bf m}^{(N)} \ket{\rm BCS} \,.
\label{8.6.25}
\end{equation}
The trial state (\ref{8.6.25}) superposes the correlated basis states
$\ket {\Psi_m^{(N)}}$ with the same amplitudes the model states 
$\ket{\Phi_m^{(N)}}$ have in the corresponding expansion of the
{\it original\/} BCS vector. It is important to note that this state 
differs from the state proposed, analyzed, and applied computationally in  
Refs.~\citenum{Fantonipairing,Fabrocinipairing,Fabrocinipairing2},
which fails to include the normalizing denominators present in
Eq.~(\ref{eq:States}).  As shown in Ref.~\citenum{HNCBCS}, this 
option leads to a meaningful gap equation only if specific diverging 
quantities are omitted.

Consider now the expectation value of an arbitrary two-body operator 
$\hat O$ with respect to the superfluid state (\ref{8.6.25}):
\begin{equation}
\left\langle\hat O\right\rangle_s  =
	\frac{{\bra {\rm CBCS}} \hat O {\ket {\rm CBCS}}}
	{\langle {\rm CBCS}\ket {\rm CBCS}} \,.
\label{8.6.26}
\end{equation}
For superfluid gaps that are small compared to the Fermi energy, it
suffices to consider the interaction of only one Cooper pair at a
time.  In that case, one need retain only the terms of {\it first order in
the deviation\/} $v^2_{\kvec} - v_{0,\kvec}^2$ and those of {\it second 
order\/} in the product $u_{\kvec}v_{\kvec}$. We refer to this as the
``decoupling approximation''.  The error introduced thereby is of
order $\varepsilon = (\Delta_F / \EF )^2$, where $\Delta_F$ is the
superfluid gap energy at the Fermi energy $\EF$.
Within this approximation, neither the pairing matrix elements nor
the single-particle energies entering the gap equation depend 
on the Bogoliubov parameters $u_{\kvec}$, $v_{\kvec}$.

The calculation of $\bigl\langle \hat H - \mu \hat N\bigr\rangle_s$ 
for correlated states is somewhat tedious\cite{CBFPairing}.  Details
may be found in Refs.~\citenum{CBFPairing,cbcs}; we only give the 
final result.  The energy of the superfluid state may be derived
from
\begin{eqnarray}
\langle \hat H - \mu \hat N \rangle_s &=& H_{oo}^{(N)} - \mu N 
+ 2 \sum_{\kvec ,\,|\, \kvec \,|\,>\KF} v_{\kvec}^2 (e_{\kvec} - \mu ) 
- 2 \sum_{\kvec , \,|\, \kvec \,|\,<\KF} u_{\kvec}^2 (e_{\kvec} - \mu ) \nonumber \\ 
&\quad& + \sum_{\kvec,\kvec'}u_\kvec v_\kvec u_{\kvec'} v_{\kvec'} 
{\cal P}_{\kvec\kvec'}
\label{5.7.13}
\end{eqnarray}
in terms of the ``pairing interaction'' specified by
\begin{eqnarray}
{\cal P}_{\kvec\kvec'} &=& {\cal W}_{\kvec\kvec'}+(|e_{\kvec}- \mu | 
+ |e_{\kvec'}- \mu |)
{\cal N}_{\kvec\kvec'}\label{eq:Pdef}\,,\\
{\cal W}_{\kvec\kvec'} &=& \bra{\kvec \uparrow ,-\kvec\downarrow}
{\cal W}(1,2)\ket{\kvec'\uparrow ,-\kvec'\downarrow}_a\,,\label{eq:Wdef}\\
{\cal N}_{\kvec\kvec'}&=&
\bra{\kvec \uparrow ,-\kvec\downarrow}
{\cal N}(1,2)\ket{\kvec'\uparrow , - \kvec'\downarrow}_a\,.
\label{eq:Ndef}\end{eqnarray}

With the result (\ref{5.7.13}), we have arrived at a formulation of
the theory which is isomorphic with the BCS theory for weakly
interacting systems. Closer inspection\cite{cbcs} reveals that our
approach corresponds to a BCS theory formulated in terms of the
scattering matrix \cite{PethickSmith}. The correlation operator $F$
serves here to tame the short-range dynamical correlations.  The effective
interaction ${\cal W}(1,2)$ is just an energy-independent
approximation of the $T$-matrix.

We may now implement the standard procedure of determining the 
Bogoliubov amplitudes $u_\kvec $, $v_\kvec $, by variation of the
energy expectation (\ref{5.7.13}).  This leads to the familiar 
gap equation
\begin{equation}
\Delta_\kvec = -\frac{1}{2}\sum_{\kvec'} {\cal P}_{\kvec\kvec'}
\frac{\Delta_{\kvec'}}{\sqrt{(e_{\kvec'}-\mu)^2 + \Delta_{\kvec'}^2}}\,.
\label{eq:gap}
\end{equation}
The conventional ({\em i.e.\/} ``uncorrelated'' or ``mean-field'') 
BCS gap equation \cite{FetterWalecka} is retrieved by replacing 
the effective interaction ${\cal P}_{\kvec\kvec'}$ by the pairing 
matrix of the bare interaction. The low-cluster-order approximations 
to the pairing interaction used by Benhar\cite{Benhar} and Pavlou 
{\em et. al.\/} \cite{Pavlou2017} are obtained by setting 
$\Gamma_{\rm dd}(r) \approx f^2(r)-1$ in Eqs.~(\ref{eq:NWloc}) 
and (\ref{eq:Weff}) and omitting the induced interaction 
$w_{\rm I}(r)$.

\section{Application to Neutron Matter}
\label{sec:results}

\subsection{Energetics}
\label{ssec:energetics}

\begin{figure}
\centerline{\includegraphics[width=0.6\textwidth,angle=-90]{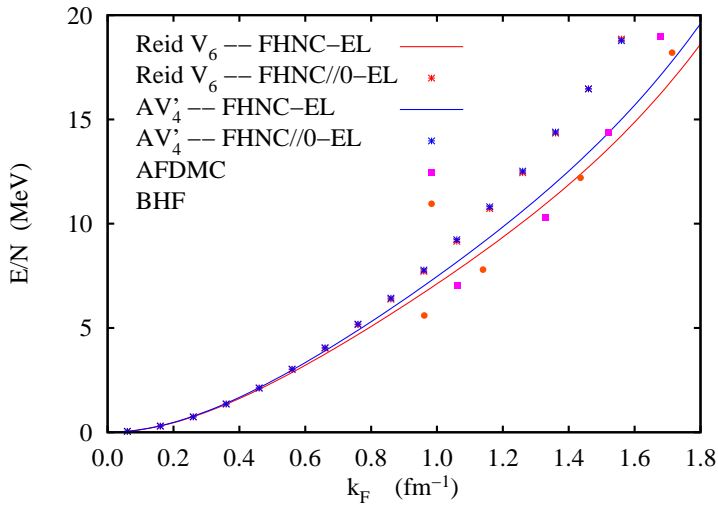}
}
\caption{(color online) Neutron-matter equation of state for the
  central component of the Reid $V_6$ soft-core potential (red line
  and stars) and for the Argonne $V_4'$ potential (blue line and
  stars), as obtained by a full FHNC-EL calculation (solid lines) and
  from the simple FHNC//0-EL approximation  (stars).  Included for
  comparison are results from the auxiliary-field diffusion Monte
  Carlo (AFDMC) method\cite{Gandolfi2009a} for the Argonne $V_{18}$
  interaction (magenta squares) and from a Brueckner-Hartree-Fock
  (BHF) calculation\cite{Baldo2012} for the Argonne $V_4'$ potential
  (orange dots).}
\label{fig:eosplot}
\end{figure}

We have carried out ground-state calculations for static properties
and superfluid pairing gaps in neutron matter based on two
representative NN interactions acting in the $T=1$ channel, namely the
central parts of the Reid soft-core potential \cite{Reid68} as formulated 
in Eqs.~(A.1)-(A.8) of Ref.~\citenum{Day81}, generally referred
to as Reid $V_6$, and the Argonne $V_4'$ potential \cite{AV18}.  In 
the density range of interest, any interaction must give very close 
to the same $E/N$ for nuclear matter and the deuteron, as long as it 
fits the $S$-wave scattering data and the deuteron.  We have carried 
out two types of calculations: Full FHNC-EL calculations as described, 
for example, in Refs.~\citenum{annals} and \citenum{polish}, and FHNC//0-EL
calculations as described in section \ref{ssec:FHNC}.  Results for the
equation of state for these two calculations, plotted as $E/N$ versus
Fermi momentum $k_F$, are shown in Fig.~\ref{fig:eosplot}.

The picture is very similar to that found for Lennard-Jones
interactions \cite{ljium}: The FHNC//0 approximation performs well up
to about half nuclear saturation density.  It is also noteworthy that
the two potentials lead to very nearly the same equation of state in
the density range considered, with the stars for the respective
FHNC//0-EL calculations overlapping.

For the Reid potential we have also examined the importance of
optimized triplet correlations ({\em i.e.,\/} non-vanishing $u_3$ in
Eq.~(\ref{eq:Jastrow}) and elementary-diagram cluster contributions
as outlined in Ref. \citenum{polish})
and found their influence negligible.  We have also tried the central
part of the full Argonne $V_{18}$ potential in the $T=1$ channel.  It
turns out that this component of the interaction is too soft to lead
to a stable solution of the Euler equation. The problem can be solved
by an artificial enhancement of the repulsive regime, but the results
depend sensitively on that enhancement factor and hence were
considered unreliable.

\subsection{BCS pairing}
\label{ssec:bcs}

Once the ground-state correlations are known, the superfluid gap 
function $\Delta_{\kvec}$ can be determined by solving the gap equation
(\ref{eq:gap}). Since we are concerned with $^1S_0$ pairing, we have 
inserted the $^1S_0$ component of the chosen potential model into 
the effective interaction (\ref{eq:Weff}).  In the phonon-exchange
correction $w_{\rm I}(r)$ the central component of the interaction is
the appropriate choice.

The gap equation was solved by the eigenvalue method with an adaptive
mesh, as outlined in the appendix of Ref.~\citenum{cbcs}.  We have
primarily adopted a free single-particle spectrum for $e_{\kvec}$ as
it occurs in Eqs.~(\ref{eq:Pdef}) and (\ref{eq:gap}).  One could also
use the actual spectrum of CBF single-particle energies
(\ref{eq:CBFspect}), calculated from the effective interactions
\cite{CBF2}, in both the pairing interaction (\ref{eq:Pdef}) and the
denominator of Eq.~(\ref{eq:gap}). We have not done this for the reason
outlined below.

\begin{figure}
\centerline{\includegraphics[width=0.6\textwidth,angle=-90]{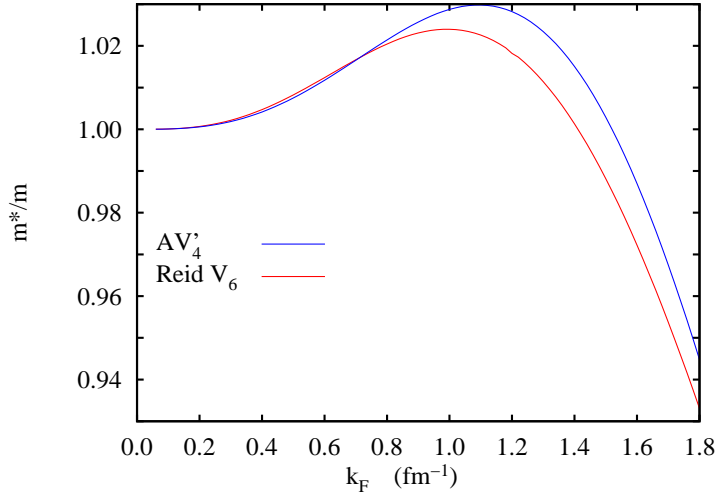}
}
\caption{(color online) Neutron effective mass for the central component 
   of the Reid $V_6$ soft-core potential (red line) and for the Argonne 
   $V_4'$ potential (blue line), as derived from the CBF single-particle 
   spectrum (\ref{eq:CBFspect}).}
\label{fig:massplot}
\end{figure}

At first glance, only the spectrum in the vicinity of the Fermi
momentum is relevant. In that regime it can be approximated quite well
in terms of an effective mass.  Fig.~\ref{fig:massplot} shows the
effective mass obtained from the CBF single-particle energies for both
potential models.
Evidently, the effective mass ratio $m^*/m$ obtained for both
potentials is very close to unity. 

However, ``first glance'' may not be sufficient; there is a subtlety
to consider: If the gap at the Fermi surface is small, we can replace
the pairing interaction $\tilde{\cal W}(k)$ by its $S$-wave matrix
element at the Fermi surface,
\begin{equation}
\tilde {\cal W}_F \equiv \frac{1}{2 \KF^2}\int_0^{2\KF} k dk \tilde{\cal W}(k)
= N{\cal W}_{\KF,\KF}\,.
\label{eq:V1S0}
\end{equation}
Then we can write the gap equation as
\begin{equation}
1 = - \tilde {\cal W}_F\int\frac{ d^3k'}{(2\pi)^3\rho}
\Bigg[\frac{1}{\sqrt{(e_{k'}-\mu)^2 + \Delta^2_{\KF}}}\label{eq:gaplowdens}
  -\frac{|e_{k'}-\mu|}{\sqrt{(e_{k'}-\mu)^2 + \Delta^2_{\KF}}}
\frac{\SF(k')}{t(k')} \Bigg]\,,\label{eq:zerorange}
\end{equation}
which is almost identical to Eq.~(16.91) in
Ref.~\citenum{PethickSmith}.  In particular, the second term, which
originates from the energy numerator generated in Eq.~(\ref{eq:gap})
by the second term of ${\cal P}_{\kvec\kvec'}$ in Eq.~(\ref{eq:Pdef}),
has the function of regularizing the integral for large $k'$. This
feature is lost if the bare mass is used in the relationship
(\ref{eq:NWloc}), and the integral (\ref{eq:gaplowdens}) diverges
unless a momentum-dependent effective mass ratio is used that
approaches unity in the limit of large momenta.

A second issue is that it has been known for a long time
\cite{BGG63,QF78} that the effective mass in nuclear systems has a
peak around the Fermi surface, however, such a peak is absent in the
CBF single-particle spectrum.  An effective-mass enhancement may be
obtained by including complex self-energy corrections; this can be
done, for example, by going to higher-order terms in CBF perturbation
theory \cite{KSJ81}.  We note that the enhancement effect is much
stronger in $^3$He (see Refs.~\citenum{ZaringhalamMass,Bengt,he3mass})
due to the softness of the spin-fluctuation mode.

\begin{figure}
\centerline{\includegraphics[width=0.6\columnwidth,angle=-90]{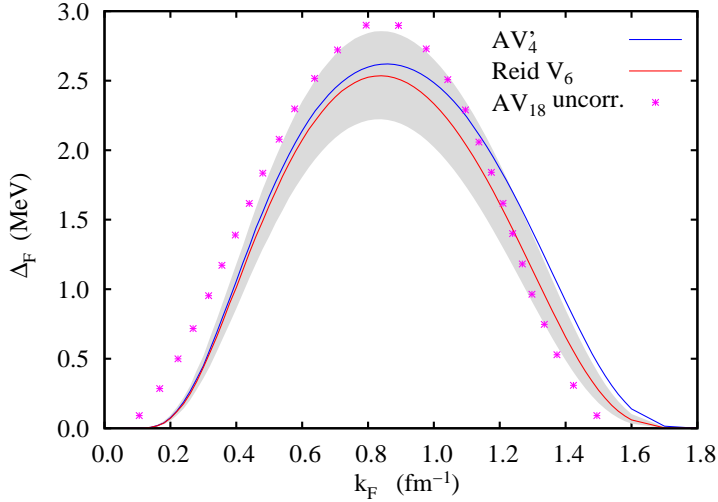}}
\caption{(color online) Superfluid gap $\Delta_{\KF}$ at the Fermi
  momentum as a function of Fermi wave number $\KF$ for the Reid $V_6$
  soft-core interaction (red curve) and the Argonne $V_4'$ potential
  (blue curve).  The gray shaded area shows the range 
  of influence an effective-mass correction can have: The lower 
  boundary of that area corresponds to $m^*/m = 0.95$ and the upper 
  boundary to $m^*/m = 1.05$. Included for comparison are results
  from a pure (``uncorrelated'') BCS gap calculation for the Argonne 
  $V_{18}$ interaction (red stars).}

\label{fig:gapplot}
\end{figure}

In view of these considerations, we have deemed it more accurate to
employ the free single-particle spectrum $e_k = t(k)$, and to study
the {\em sensitivity\/} of our results to changes in the effective
mass. Our results for the superfluid gap for the two potentials are
shown in Fig.~\ref{fig:gapplot}. Evidently the difference of the gap
between these two potential models is almost negligible and certainly
within the accuracy of the FHNC approximations. To determine the
importance of effective-mass corrections we have also solved the gap
equation assuming effective-mass ratios between $m^*/m = 0.95$ and
$m^*/m = 1.05$ in both the pairing interaction (\ref{eq:Pdef}) and the
energy denominator (\ref{eq:eden}). The results for the gap define the
gray area in Fig.~\ref{fig:gapplot}; their spread provides a
conservative estimate of the importance of a non-trivial
single-particle spectrum.

\subsection{Consequences for many--body theory}
\label{ssec:MBT}
To conclude this section, let us look more closely at different aspects
of the convergence of cluster expansion and resummation techniques.
Apart from cold atomic gases -- which with rare exceptions like the
``unitary limit'' pose no challenges to modern many-body theory --
pure neutron matter at subnuclear and nuclear densities is, apart from  the
complications introduced by the nucleon-nucleon force, one of the most 
lenient many-particle systems provided by nature.  This is largely due 
to the low density of the system, as measured for example by the ratio 
of the pion Compton wavelength \cite{Reid68}, $\lambda_\pi = 1/\mu_\pi 
= \hbar/m_\pi c \approx 1.4~{\rm fm}$, or the radius of the ``hard 
core,'' $\sigma \sim 0.7-0.9~{\rm fm}$, to the average particle 
spacing at the given density $\rho = k_F^3/3\pi^2$.  Thus, at $\KF 
= 1.4~{\rm fm}^{-1}$ the density is $\rho\approx 0.06\sigma^{-3}$, 
which corresponds to only 20 percent of the saturation density of 
$^3$He.

Evidence for the good convergence of many-body theory for
neutron matter in the density regime relevant for $^1S_0$ pairing 
is already provided in Fig.~\ref{fig:eosplot}, which shows that the 
very simple FHNC//0 approximation for the energy is quite accurate.
In fact, even the very simple two-body cluster 
approximation
\begin{equation}
  \left(\frac{E}{N}\right)_2 = \frac{T_F}{N}
  + \frac{\rho}{2}\int d^3r
  \left[\left(1+ \Gamma_{\!\rm dd}(r)\right)v(r)
    +  \frac{\hbar^2}{m}\left|\nabla\sqrt{1+\Gamma_{\!\rm dd}(r)}\right|^2
    \right]g_F(r),
  \label{eq:etwo}
  \end{equation}
in which $T_F$ is the kinetic energy of the free Fermi gas and $g_F(r)
= 1 - \ell^2(r\KF)/2$ its pair distribution function, yields results
virtually identical to the FHNC//0 results plotted in
Fig.~\ref{fig:eosplot}.  We have refrained from showing these results
in order not to obscure the figure.  Note that one can of course
identify $1+ \Gamma_{\!\rm dd}(r)$ with $f^2(r)$ in
Eq.~(\ref{eq:etwo}).

\begin{figure}
\centerline{\includegraphics[width=0.6\columnwidth,angle=-90]{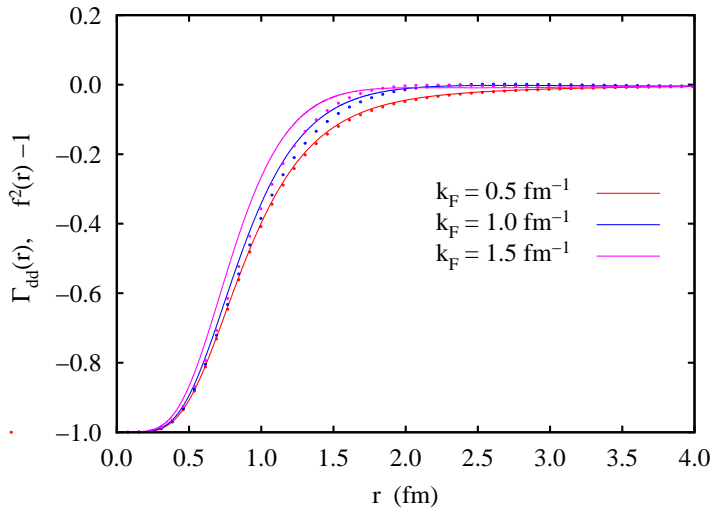}
}
\caption{(color online) Plots of the ``dressed'' correlation
  function $\Gamma_{\rm dd}(r)$ (solid lines) for three representative
  densities, as indicated in the legend. Also shown is the pair 
  correlation function $f^2(r)-1$ (dotted lines). Note that this
  function is calculated {\em a posteriori\/} from the solution of the
  Euler equation; the generic many-body method spelled out in Section
  \ref{ssec:FHNC} never needs to introduce this quantity.}
\label{fig:gammaplot}
\end{figure}

These findings are consistent with the fact that the optimal results
for $1+ \Gamma_{\!\rm dd}(r)$ and $f^2(r)$ are not very different. To
demonstrate this, both functions are plotted in Fig.~\ref{fig:gammaplot} 
for three representative densities, $\KF = 0.5,\, 1.0,\,$ and
$1.5~{fm^{-1}}$. At the lowest density, the two functions are
practically identical. As the density increases, $\Gamma_{\!\rm dd}(r)$
becomes slightly steeper in the attractive regime of the interaction.

From these results, one might be led to conclude that
low-order methods are also adequate for calculating the superfluid
gap. We remind the reader, however, of the discussions in
Sect.~\ref{ssec:sensitivity} on the both the sensitivity of
quantities other than the energy to the correlation functions, and
to the convergence rate of cluster expansions.  Accordingly, we
have examined the consequences of two approximations: Leaving out
the energy numerator generated by the pairing interaction 
(\ref{eq:Pdef}) and leaving out the induced interaction $w_{\rm I}(r)$.

\begin{figure}
\centerline{\includegraphics[width=0.6\columnwidth,angle=-90]{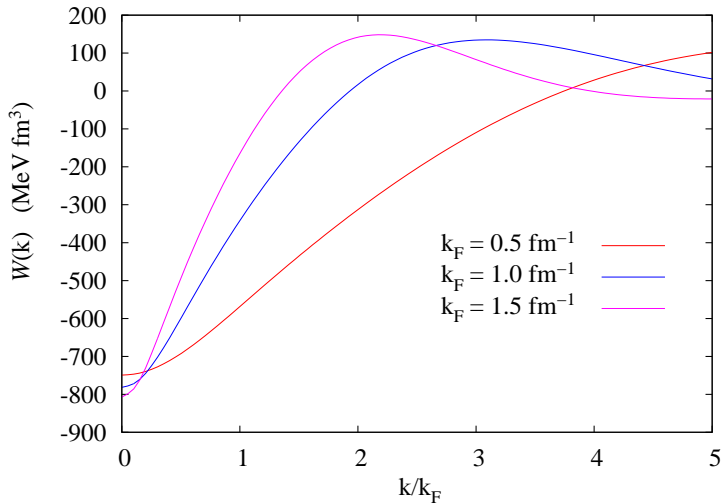}
}
\caption{(color online) Momentum dependence of the pairing interaction 
${\cal W}(k) \equiv \tilde{\cal W}(k)/\rho$ for three representative
densities as indicated in the legend.}
\label{fig:potplot}
\end{figure}

The most important function of the energy numerator is to regularize
the integral in the gap equation for contact interactions, as witnessed
in Eq.~(\ref{eq:gaplowdens}). The situation being discussed at that point 
is, of course, extreme.  More generally, one would expect that the energy
numerator term is important whenever the pairing interaction
$\tilde{\cal W}(k)$ does not fall off sufficiently rapidly for large
momenta. This is indeed the case: We show in Fig. \ref{fig:potplot}
the interaction $\tilde{\cal W}(k)$ for three representative
densities. Evidently, the pairing interaction does not fall off
rapidly above $\KF$. The effect is, of course, most pronounced for
low densities. Although the gap is determined solely by the pairing
matrix element $\tilde{\cal W}_F$ in the limit of an infinitesimal
gap, one must expect significant finite-range effects in the
present case where the gap is of the order of 10 to 50 percent of
the Fermi energy.
  
The second new aspect is the appearance of the induced interaction
term $w_{\rm I}(r)$ appearing in the pairing interaction (cf.\
Eqs.~(\ref{eq:Weff}) and (\ref{eq:inducedFermi0}). This term describes
the exchange of particle-hole excitations \cite{parquet1} and is one
of the important effects introduced into the CBF version of BCS
theory. While the gap equation includes the summation of ladder
diagrams\cite{CMS,NozieresSchmittRink} and can, at least in principle,
deal with bare hard-core interactions, the particle-hole reducible
diagrams described by $w_{\rm I}(r)$ introduce new physics.

Ignoring the induced interaction $w_{\rm I}(r)$ leads to the
two-body approximation
\begin{equation}
  {\cal W}_2(r) = \left[1+\Gamma_{\rm dd}(r)\right]v(r) + 
\frac{\hbar^2}{m}\left|\nabla\sqrt{1+\Gamma_{\!\rm dd}(r)}\right|^2
\label{eq:Wtwo}
\end{equation}
for the pairing interaction.  We note that in this case one can again 
identify $\Gamma_{\rm dd}(r)\Rightarrow f^2(r)-1$.

\begin{figure} \centerline{\includegraphics[width=0.6\columnwidth,angle=-90]{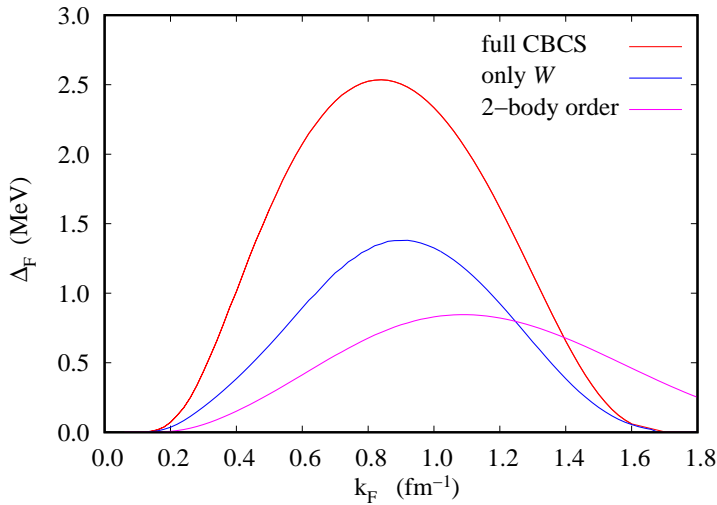}
}
\caption{(color online) This figure shows the consequences of the two
  approximations discussed in the text for the magnitude of the gap at
  the Fermi surface. The curve ``full CBCS'' (red) is identical to 
  that shown in Fig.~\ref{fig:gapplot}; the curve ``only ${\cal W}$''
  (blue) shows the consequence of omitting the energy-numerator term 
  generated by the CBF pairing interaction, and the curve ``2-body
  order'' (magenta) is obtained by using the two-body approximation 
  (\ref{eq:Wtwo}) while also leaving out the energy-numerator term. In 
  this last case, inclusion of the energy-numerator term does not lead 
  to sensible results because the cancellation illustrated by 
  Eq.~(\ref{eq:gaplowdens}) is violated.}

\label{fig:manygaps}
\end{figure}

Fig.~\ref{fig:manygaps} demonstrates the impact on the calculated 
energy gap of the two approximations identified above, for the case of
the Reid potential.  Evidently, both simplifications have rather dramatic 
effects, being enhanced by the nominally exponential dependence of 
the gap on the pairing interaction.  At this point we are not prepared 
to describe or affirm any systematics of the effects of these approximations.
However, the close proximity of the ``full CBCS'' results
and those for the bare interaction shown in Fig.~\ref{fig:gapplot}
would seem to be coincidental, stemming from competing corrections.

\subsection{Comparison with Previous Gap Calculations}
\label{ssec:comparison}
The work we report represents the most rigorous calculation yet performed
for nuclear systems within correlated BCS theory.  It is therefore 
of special interest to compare its results with those of earlier
calculations of the $^1S_0$ pairing gap for neutron matter based on 
microscopic many-body theories, where meaningful conclusions might 
be drawn.  

Informative comparison of the predictions of previous gap calculations 
-- as represented for example by the aforementioned summary figure 
in the review by Gezerlis {\em et al.}\cite{Gezerlis2014} -- is rendered 
problematic by the diversity of methods applied, interactions adopted,
and assumptions made ({\em e.g.,} whether or not polarization effects 
from exchange of density and/or spin-density fluctuations are 
included).  Nevertheless, some specific and non-specific comparisons 
may be useful.  

Fig.~\ref{fig:manygaps} includes data plotted for a pure-BCS
calculation in which the pairing interaction is the bare potential in
the $^1S_0$ channel of the Argonne $V_{18}$ interaction, used along
with free single-particle $e_k$.  The BCS result for Argonne $V_{18}$,
calculated by the separation method of Ref.~\citenum{KKC96}, was taken
from Ref.~\citenum{Yuan}.  (For this present purpose, the distinction
between the original Argonne $V_{18}$ interaction and Argonne $V_4'$
should be immaterial.)  Corresponding bare-BCS results for the Reid
$V_6$ choice (displayed in Ref.~\citenum{KKC96} but not plotted here)
are very close to those shown for Argonne $V_{18}$, as expected.  What
is unexpected is that our CBF results for the Argonne case show only a
modest suppression (about 15\%) of the $\Delta_F$ maximum, which
occurs slightly above $0.8~{\rm fm}^{-1}$ in both calculations.  The
approximately Gaussian shape of $\Delta_F$ vs.\ $k_F$ shifts to
slightly lower $k_F$ in the absence of Jastrow-Feenberg correlations.
It is obvious from Fig.~\ref{fig:manygaps} that this near concurrence
cannot be attributed to unimportance of the correlations introduced in
the CBF treatment.  It is possible that this feature is due in part to
the presence of the induced-interaction term $w_{\rm I}(r)$ in the
effective pairing interaction coming from density fluctuations, which
is expected to enhance the gap relative to that given by direct part
of ${\cal W}(r)$.

The most recent microscopic calculations of the $^1S_0$ gap incorporating 
Jastrow-Feenberg two-body correlations are those of Pavlou 
{\em et~al.}\cite{Pavlou2017}, described briefly in Sec.~\ref{ssec:sensitivity}.
Their variational CBF study was carried out for each of two different 
parametrized forms the Jastrow factor $f(r)$, subject to a constraint on 
its ``wound,'' as outlined briefly in Sec.~\ref{ssec:sensitivity}.  Both 
of these forms have been used in earlier work: one, referred to as the 
``Benhar'' choice, has one free parameter but allows $f(r)$ to overshoot 
unity, whereas the ``Dav\'e'' choice has two free parameters but no 
overshoot.  Restricted minimization was performed on an approximation 
to the energy expectation $E/N$ that retains only the leading (zeroth) 
order of its cluster expansion, neglecting terms of first and higher 
orders in a dimensionless small parameter $\xi$ that grows with 
density. (While its value remains well below 0.05 in the relevant 
density regime, the implied rate of convergence of the expansion of 
$E/N$ {\it does not\/} extend to approximants of $\Delta_F$.)  

The approach adopted in Ref.~\citenum{Pavlou2017} may be considered the 
simplest implementation of correlated CBF theory. The effective pairing 
interaction it generates differs from its FHNC-EL counterparts 
in two essential respects: it lacks precisely those ingredients that 
are the subjects of the above discussion of the ``many-body consequences'' 
of our work based on FHNC-EL theory, namely the energy numerator 
term and the induced interaction entering the effective interaction
$\cal W$.

This same statement applies to the variational component of the 
correlated BCS approach applied much earlier by Chen {\em et al.}\cite{JWCgap}, 
in which the $^1S_0$ gap in neutron matter was estimated based on the 
central, $V_4$ part of the Reid $V_6$ soft-core interaction.  In that case 
$\Delta_F$ was found to peak at about $0.75~{\rm fm}^{-1}$ with a maximum 
value close to 3.2 MeV, a result based in fact on the Dav\'e form for 
$f(r)$.  It should then be no surprise that with negligible differences, 
Pavlou {\em et al.} obtained to almost exactly the same result for $\Delta_F$ 
versus $k_F$, although the Argonne $V_{18}$ interaction was assumed.  
It should be said that all of the tests we have made support the 
assertion that, when the $^1S_0$ gap is calculated by the same method 
with different inputs for the bare NN interactions but otherwise the 
same assumptions ({\em e.g.,} for the single particle energies $e_k$), 
virtually identical results will be obtained $\Delta_F$, provided 
the NN interaction chosen reproduces the NN scattering data up to 
laboratory energies relevant for $k_F$ below about $1.5~{\rm fm}^{-1}$.  
Indeed, this a well established property for the BCS gap\cite{H-J1998}.

As already pointed out, the maximum gap value obtained by Pavlou {\em
  et al.}\ with AV$_{18}$ for their two optimized correlation
functions differ by nearly a factor two (3.3 MeV for the Dav\'e form
at $k_F = 0.85~{\rm fm}^{-1}$ and 1.8 MeV at $k_F = 0.75~{\rm
  fm}^{-1}$ for the Benhar choice) -- reflecting the extreme
sensitivity of the gap to inputs for the effective interaction.
Recognizing that the induced interaction and energy-numerator terms
are absent in these two calculations, the information provided by
Fig.~\ref{fig:manygaps} on the relative contributions of these
additional terms suggests that the results obtained in
Ref.~\citenum{Pavlou2017} for the Benhar correlation function are to
be favored over those for the Dav\'e form.

Turning to other microscopic calculations designed to provide accurate
predictions for the $^1S_0$ gap in neutron matter, we first single out
the study of Cao, Lombardo, and Schuck \cite{Cao2006}, carried out
within the framework of Brueckner theory.  Mean-field theory for the
superfluid state, as represented by the pure-BCS treatment, was
modified by replacement of the bare pairing interaction with a proper
vertex part, which includes an induced interaction describing the
competition between the attractive density excitations and their
repulsive spin-density counterparts ({\em i.e.,\/} screening or polarization
corrections).  In-medium corrections were also introduced into the
self-energies $e_k$, corresponding to both dispersion and
Fermi-surface depletion.  The quenching of the gap due to exchange of
spin-density fluctuations was found to be less extreme than indicated
by some previous studies.  Results based on a free single-particle
spectrum were also reported, allowing more direct comparison with our
results.  For the free spectrum, Cao {\rm et al.}\cite{Cao2006} find a
maximum $\Delta_F$ of about 2.7 MeV occurring close to $k_F =
0.85~{\rm fm}^{-1}$, based on the Argonne $V_{18}$ interaction.  The
close agreement with our CBCS results shown in Fig.~\ref{fig:gapplot}
is remarkable, but provocative, as our treatment does not include an
induced interaction term corresponding to spin-density fluctuations.
On the other hand, the treatment of screening effects in the two
approaches is not directly comparable.  For a recent intensive
computational analysis of medium polarization in asymmetric nuclear
matter, see Ref.~\citenum{Zhang2016}.

The auxiliary-field diffusion Monte Carlo algorithm (AFDMC) purports to 
yield accurate results for pairing gaps in neutron matter and other 
many-fermion systems \cite{Gandolfi2009a,Gandolfi2009b}.  For the 
present considerations, this algorithm has two distinctive features. 
\begin{itemize}
\item[(i)] Unlike the BCS state, the trial wave function $\Psi_T$ 
that is propagated in imaginary time describes a definite number of 
particles $N$, even or odd.  The part of $\Psi_T$ that describes 
pairing is essentially the projection of the BCS state onto the 
$N$-particle Hilbert space, which is a Pfaffian.  Correlations are 
otherwise introduced into $\Psi_T$ by a Jastrow factor.
\item[(ii)]
The pairing gap is constructed as a difference of energies obtained
for different particle numbers, 
\end{itemize}
\begin{equation}
\Delta = E(N) - \frac{1}{2} \left[E(N+1) + E(N-1)\right].
\end{equation}
Whatever the merits and deficits of this approach, they are generally 
different from those of traditional many-body theory; consequently,
AFDMC tends to be regarded as an essentially independent arbiter in 
judging the quality of such traditional method when comparison
can be made.  The number of data points shown in 
Ref.~\citenum{Gandolfi2009b} for the $^1S_0$ gap in neutron
matter do not allow a precise identification of the peak value
of $\Delta_F$, but it lies slightly above 2 MeV, reached slightly
above $0.6~{\rm fm}^{-1}$.  The error bars shown are roughly half an
MeV.  We present this result only to provide a balanced perspective 
on the current status of the problem, but restrain from drawing 
any conclusions about its bearing on the quality of our calculations.
  
\section{Summary and Prospects}
\label{sec:summary}

In this paper we have described new calculations of the pairing gap 
in the $^1S_0$ partial-wave channel. Our findings have been analyzed
and discussed in the preceding section.

The most interesting result of previous\cite{cbcs} work along these
lines is the appearance of a divergence of solutions of the FHNC-EL 
equations that occurs, as a function of potential strength, well 
before the divergence of the vacuum scattering length $\a0$ of the 
interaction potential. This divergence of solutions of the FHNC-EL 
equations is analogous to the spinodal instability often discussed in 
earlier literature, with the principled and practical conclusion that the 
FHNC-EL equations for the {\em homogeneous\/} system have no solutions 
if $F_0^s < -1$, {\em i.e.\/} if the system is unstable in the 
particle-hole channel. In Ref.~\citenum{cbcs}, divergence of the FHNC-EL 
equations in the case of a diverging in-medium scattering length gave 
evidence that the ground state is unstable against dimerization.  The 
appearance of such instabilities whenever the assumptions on the state 
of the system fail -- here, assumption of a non-dimerized phase; in the case 
of particle-hole instabilities, of a uniform system -- is a unique feature 
of theories such as FHNC-EL that enjoy the topological completeness 
of parquet diagrams.

In the calculations being reported, we have not encountered such an
instability, which could be taken as evidence that medium-driven
formation of dineutrons in low-density neutron matter does not occur,
or, in current terminology, that a BEC-BCS crossover
\cite{MargueronPRC76,PhysRevC.94.034004,NuclBCSBEC,bel}
does not
take place. This is remarkable in view of the fact that, at low
densities ($\KF\approx 0.2\,{\rm fm}^{-1}$), the gap reaches 0.45
times the Fermi energy $\EF$ which is comparable to what is found in
the unitary Fermi gas at the BCS-BEC
crossover\cite{annurev-conmatphys-031113-133829}. It should be
mentioned that a recent study \cite{Stein} of the phase
diagram of spin-polarized neutron matter revealed signatures that can
be interpreted \cite{NozieresSchmittRink} as a precursor of such a
crossover.

There are four areas where the present calculation can be improved:

(a) As pointed out above, the FHNC-EL method sums all ring and ladder 
diagrams. It does that, however, in a ``collective approximation'' 
of the particle-hole and the particle-particle propagators \cite{polish} 
that treats the correlations between particles within the Fermi sea 
in an average way. Since pairing occurs between particles at the 
Fermi surface, it must be examined to what extent the average treatment 
of correlations is appropriate. The route to improve upon this aspect is 
well charted within CBF theory, and earlier studies \cite{shores,CCKS86} 
have demonstrated that CBF corrections to the pairing matrix elements 
can indeed be significant.

(b) Related to (a): whereas the effect of density fluctuations (exchange 
of virtual phonons) has been included in the CBF pairing interaction
in an average-propagator sense, effects of spin-density fluctuations
are not taken into account.  Based on Landau parameters and some
microscopic efforts \cite{JWCgap,CKY76,Wam93,Sch96}, density fluctuations 
produce a modest enhancement of the pairing gap, whereas the spin-density
channel generates a dominant suppression. Without introducing explicit
spin-dependent correlations into the basis functions of the CBF treatment, 
their perturbative treatment within the CBF framework would be required.

(c) In the present work, in-medium effects on the self-energy input 
$e_k$ to the gap equation have not been pursued quantiatively.
This shortcoming warrants further attention in subsequent applications
of correlated BCS theory.
 
(d) The most severe approximation made in this work is the use of
state-independent correlation functions, albeit the two-nucleon
interaction is exquisitely state-dependent. Introduction of a 
correlation operator $F$ in Eq.~(\ref{eq:wavefunction}) that 
contains spin-, isospin-, tensor-, and more complicated operators 
in the two-body correlation vehicle $u_2(ij)$ figuratively opens 
Pandora's box.  This complexity has been largely dealt with in 
rather simple approximations that either completely omit commutator 
terms \cite{FantoniSpins,Wiri78} or in a ``single-operator-chain'' 
approximation \cite{PAW79}, which only sums the ring diagrams of 
state-dependent correlations.  Unfortunately, for modern nucleon-nucleon 
interactions, which may have different core sizes in the singlet and 
triplet channels, the contributions of commutator diagrams can 
be huge \cite{SpinTwist}.  It remains to be seen how important 
these effects are in the problem considered here, but at higher 
densities they can be decisive.

\begin{acknowledgements}

This work was supported, in part, by the College of Arts and
Sciences, University at Buffalo SUNY, and the Austrian Science Fund
project I602 (to EK).  JWC acknowledges support from the McDonnell
Center for the Space Sciences, and expresses gratitude to the
University of Madeira and its Center for Mathematical Sciences
for gracious hospitality during periods of extended residence.

\end{acknowledgements}
\pagebreak


\begin{thebibliography}{10}
\providecommand{\url}[1]{{#1}}
\providecommand{\urlprefix}{URL }
\expandafter\ifx\csname urlstyle\endcsname\relax
  \providecommand{\doi}[1]{DOI \discretionary{}{}{}#1}\else
  \providecommand{\doi}{DOI \discretionary{}{}{}\begingroup
  \urlstyle{rm}\Url}\fi

\bibitem{BCS}
J.~Bardeen, L.N. Cooper, J.R. Schrieffer, Phys. Rev. \textbf{108}, 1175 (1957)

\bibitem{BCS50book}
R.~Broglia, V.~Zelevensky, \emph{Fifty Years of Nuclear BCS} (World Scientific,
  Singapore, 2013)

\bibitem{BMP}
A.~Bohr, B.R. Mottelson, D.~Pines, Phys. Rev. \textbf{110}, 936 (1958)

\bibitem{CMS}
L.N. Cooper, R.L. Mills, A.M. Sessler, Phys. Rev. \textbf{114}, 1377 (1959)

\bibitem{Schrieffer1999}
J.R. Schrieffer, \emph{Theory Of Superconductivity (Advanced Books Classics)},
  revised edn. (Perseus Books, 1999)

\bibitem{Mills}
R.L. Mills, A.M. Sessler, S.A. Moszkowski, D.G. Shankland, Phys. Rev. Lett.
  \textbf{3}, 381 (1959)

\bibitem{Reid68}
R.V. {Reid, Jr.}, Ann. Phys. (NY) \textbf{50}, 411 (1968)

\bibitem{JWCgap}
J.M.C. Chen, J.W. Clark, R.D. Dav{\'e}, V.V. Khodel, Nucl. Phys. A
  \textbf{555}, 59 (1993)

\bibitem{KKC96}
V.A. Khodel, V.V. Khodel, J.W. Clark, Nucl. Phys. A \textbf{598}, 390 (1996)

\bibitem{YangNC}
J.W. Clark, C.H. Yang, Lett. Nuovo Cimento \textbf{3}, 272 (1970)

\bibitem{YangClarkBCS}
C.H. Yang, J.W. Clark, Nucl. Phys. A \textbf{174}, 49 (1971)

\bibitem{YangThesis}
C.H. Yang, Theory of superfluidity in strongly-interacting many fermion systems
  with applications to pure neutron matter and symmetrical nuclear matter.
\newblock Ph.D. thesis, Washington University (1971)

\bibitem{Johnreview}
J.W. Clark, in \emph{Progress in Particle and Nuclear Physics}, vol.~2, ed. by
  D.H. Wilkinson (Pergamon Press Ltd., Oxford, 1979), pp. 89--199

\bibitem{Chao}
N.C. Chao, J.W. Clark, C.H. Yang, Nucl. Phys. A \textbf{179}, 320 (1972)

\bibitem{KroTrieste}
E.~Krotscheck, in \emph{Introduction to Modern Methods of {Q}uantum Many--Body
  Theory and their Applications}, \emph{Advances in {Q}uantum Many--Body
  Theory}, vol.~7, ed. by A.~Fabrocini, S.~Fantoni, E.~Krotscheck (World
  Scientific, Singapore, 2002), pp. 267--330

\bibitem{CBFPairing}
E.~Krotscheck, R.A. Smith, A.D. Jackson, Phys. Rev. B \textbf{24}, 6404 (1981)

\bibitem{HNCBCS}
E.~Krotscheck, J.W. Clark, Nucl. Phys. A \textbf{333}, 77 (1980)

\bibitem{cbcs}
H.H. Fan, E.~Krotscheck, T.~Lichtenegger, D.~Mateo, R.E. Zillich, Phys. Rev. A
  \textbf{92}, 023640 (2015)

\bibitem{mag7}
J.W. Clark, A.~Sedrakian, M.~Stein, X.G. Huang, V.A. Khodel, V.R. Shaginyan,
  M.V. Zverev, Journal of Physics: Conference Series \textbf{702}, 012012
  (2016)

\bibitem{Gorkov}
L.~Gorkov, T.K. Melik-Barkhudarov, Sov. Phys. JETP \textbf{13}, 1018 (1961)

\bibitem{heiselbergPRL00}
H.~Heiselberg, C.J. Pethick, H.~Smith, L.~Viverit, Phys. Rev. Lett.
  \textbf{85}, 2418 (2000)

\bibitem{Pavlou2017}
G.E. Pavlou, E.~Mavrommatis, C.~Moustakidis, J.W. Clark, Eur. Phys. J. A
  \textbf{53}, 96 (2017)

\bibitem{AV18}
R.B. Wiringa, V.G.J. Stoks, R.~Schiavilla, Phys. Rev. C \textbf{51}, 38 (1995)

\bibitem{Gezerlis2014}
A.~Gezerlis, C.J. Pethick, A.~Schwenk, in \emph{Novel Superfluids}, vol.~2, ed.
  by K.H. Bennemann, J.B. Ketterson (Oxford University Press, 2014), chap.~22

\bibitem{Wiringa2002}
R.B. Wiringa, S.C. Pieper, Phys. Rev. Lett. \textbf{89}, 182501 (2002)

\bibitem{50yrs}
J.W. Clark, in \emph{Fifty Years of Nuclear BCS}, ed. by R.A. Broglia,
  V.~Zelevinsky (World Scientific, Singapore, 2013), chap.~27, pp. 360--376

\bibitem{parquet1}
A.D. Jackson, A.~Lande, R.A. Smith, Physics Reports \textbf{86}(2), 55 (1982)

\bibitem{parquet2}
A.D. Jackson, A.~Lande, R.A. Smith, Phys. Rev. Lett. \textbf{54}, 1469 (1985)

\bibitem{BishopValencia}
R.F. Bishop, in \emph{Condensed Matter Theories}, vol.~10, ed. by M.~Casas,
  J.~Navarro, A.~Polls (Nova Science Publishers, Commack, New York, 1995),
  vol.~10, pp. 483--508

\bibitem{PairDFT}
E.~Krotscheck, Phys. Lett. A \textbf{190}, 201 (1994)

\bibitem{FeenbergBook}
E.~Feenberg, \emph{Theory of {Q}uantum Fluids} (Academic, New York, 1969)

\bibitem{polish}
E.~Krotscheck, J. Low Temp. Phys. \textbf{119}, 103 (2000)

\bibitem{ljium}
J.~Egger, E.~Krotscheck, R.E. Zillich, J. Low Temp. Phys. \textbf{165}, 275
  (2011)

\bibitem{annals}
E.~Krotscheck, Ann. Phys. (NY) \textbf{155}, 1 (1984)

\bibitem{CBF2}
E.~Krotscheck, J.W. Clark, Nucl. Phys. A \textbf{328}, 73 (1979)

\bibitem{BeliaevLesHouches}
S.T. Beliaev, in \emph{Lecture Notes of the 1957 Les Houches Summer School},
  ed. by C.~DeWitt, P.~Nozi{\`e}res (Dunod, 1959), pp. 343--374

\bibitem{Fantonipairing}
S.~Fantoni, Nucl. Phys. A \textbf{363}, 381 (1981)

\bibitem{Fabrocinipairing}
A.~Fabrocini, S.~Fantoni, A.Y. Illarionov, K.E. Schmidt, Phys. Rev. Lett.
  \textbf{95}, 192501 (2005)

\bibitem{Fabrocinipairing2}
A.~Fabrocini, S.~Fantoni, A.Y. Illarionov, K.E. Schmidt, Nucl. Phys. A
  \textbf{803}, 137 (2008)

\bibitem{PethickSmith}
C.J. Pethick, H.~Smith, \emph{{B}ose-{E}instein Condensation in Dilute Gases},
  second edition edn. (Cambridge University Press, Cambridge, UK, 2008)

\bibitem{FetterWalecka}
A.L. Fetter, J.D. Walecka, \emph{{Q}uantum Theory of Many-Particle Systems}
  (McGraw-Hill, New York, 1971)

\bibitem{Benhar}
O.~Benhar, G.D. Rosi, G.~Salvi, J. Low Temp. Phys.  (2017).
\newblock (this volume, arXiv:1305.4659)

\bibitem{Day81}
B.D. Day, Phys. Rev. C \textbf{24}, 1203 (1981)

\bibitem{Gandolfi2009a}
S.~Gandolfi, A.~Illarionov, K.E. Schmidt, F.~Pederiva, S.~Fantoni, Phys. Rev. C
  \textbf{79}, 054005 (2009)

\bibitem{Baldo2012}
M.~Baldo, A.~Polls, A.~Rios, H.J. Schulze, I.~Vida\~na, Phys. Rev. C
  \textbf{86}, 064001 (2012)

\bibitem{BGG63}
G.~Brown, J.~Gunn, P.~Gould, Nucl. Phys. \textbf{46}, 598 (1963)

\bibitem{QF78}
P.~Quentin, H.~Flocard, Annual Review of Nuclear and Particle Science
  \textbf{28}, 523 (1978)

\bibitem{KSJ81}
E.~Krotscheck, R.A. Smith, A.D. Jackson, Phys. Lett. B \textbf{104}, 421 (1981)

\bibitem{ZaringhalamMass}
G.E. Brown, C.J. Pethick, A.~Zaringhalam, J. Low Temp. Phys. \textbf{48}, 349
  (1982)

\bibitem{Bengt}
B.L. Friman, E.~Krotscheck, Phys. Rev. Lett. \textbf{49}, 1705 (1982)

\bibitem{he3mass}
E.~Krotscheck, J.~Springer, J. Low Temp. Phys. \textbf{132}, 281 (2003)

\bibitem{NozieresSchmittRink}
P.~Nozi{\'e}res, S.~Schmitt-Rink, J. Low Temp. Phys. \textbf{59}, 195 (1985)

\bibitem{Yuan}
L.~Yuan, Three-body pairing interaction effect on superfluidity with
  applications to neutron star matter.
\newblock Ph.D. thesis, Washington University (2011)

\bibitem{H-J1998}
{\O}.~Elgar{\o}y, M.~Hjorth-Jensen, Phys. Rev. C \textbf{57}, 1174 (1998)

\bibitem{Cao2006}
L.G. Cao, U.~Lombardo, P.~Schuck, Phys. Rev. C \textbf{74}, 064301 (2006)

\bibitem{Zhang2016}
S.S. Zhang, L.G. Cao, U.~Lombardo, P.~Schuck, Phys. Rev. C \textbf{93}, 044329
  (2016)

\bibitem{Gandolfi2009b}
S.~Gandolfi, A.~Illarionov, F.~Pederiva, K.E. Schmidt, S.~Fantoni, Phys. Rev. C
  \textbf{80}, 045802 (2009)

\bibitem{MargueronPRC76}
J.~Margueron, H.~Sagawa, K.~Hagino, Phys. Rev. C \textbf{76}, 064316 (2007)

\bibitem{PhysRevC.94.034004}
F.~Isaule, H.F. Arellano, A.~Rios, Phys. Rev. C \textbf{94}, 034004 (2016)

\bibitem{NuclBCSBEC}
M.~Stein, A.~Sedrakian, X.G. Huang, J.W. Clark, Phys. Rev. C \textbf{90},
  065804 (2014)

\bibitem{bel}
V.A. Khodel, J.W. Clark, V.R. Shaginyan, M.V. Zverev, Physics of Atomic Nuclei
  \textbf{77}, 1145 (2014)

\bibitem{annurev-conmatphys-031113-133829}
M.~Randeria, E.~Taylor, Annual Review of Condensed Matter Physics
  \textbf{5}(1), 209 (2014)

\bibitem{Stein}
M.~Stein, A.~Sedrakian, X.G. Huang, J.W. Clark, Phys. Rev. C \textbf{93},
  015802 (2016)

\bibitem{shores}
A.D. Jackson, E.~Krotscheck, D.~Meltzer, R.A. Smith, Nucl. Phys. A
  \textbf{386}, 125 (1982)

\bibitem{CCKS86}
J.M.C. Chen, J.W. Clark, E.~Krotscheck, R.A. Smith, Nucl. Phys. A \textbf{451},
  509 (1986)

\bibitem{CKY76}
J.W. Clark, C.~G.K{\"a}llman, C.H. Yang, D.A. Chakkalakal, Phys. Lett. B
  \textbf{61}(4), 331 (1976)

\bibitem{Wam93}
J.~Wambach, T.~Ainsworth, D.~Pines, Nucl. Phys. A \textbf{555}, 128 (1993)

\bibitem{Sch96}
H.J. Schulze, J.~Cugnon, A.~Lejeune, M.~Baldo, U.~Lombardo, Phys. Lett. B
  \textbf{375}, 1 (1996)

\bibitem{FantoniSpins}
S.~Fantoni, S.~Rosati, Nuovo Cimento \textbf{43A}, 413 (1977)

\bibitem{Wiri78}
R.B. Wiringa, V.R. Pandharipande, Nucl. Phys. A \textbf{299}, 1 (1978)

\bibitem{PAW79}
V.R. Pandharipande, R.B. Wiringa, Rev. Mod. Phys. \textbf{51}(4), 821 (1979)

\bibitem{SpinTwist}
E.~Krotscheck, Nucl. Phys. A \textbf{482}, 617 (1988)

\end{thebibliography}

\bibliographystyle{spphys.bst}
\end{document}